\documentclass[twoside,twocolumn,english,superscriptaddress,nofootinbib,preprintnumbers, aps]{revtex4}
\usepackage{color}
\usepackage{amsmath}
\usepackage{graphicx}
\usepackage{amssymb}
\usepackage{esint}
\usepackage[hyperfootnotes=true,unicode=true, 
 bookmarks=true,bookmarksnumbered=false,bookmarksopen=false,
 breaklinks=false,pdfborder={0 0 1},backref=false,colorlinks=true]
 {hyperref}

\makeatletter

\@ifundefined{textcolor}{}
{%
 \definecolor{BLACK}{gray}{0}
 \definecolor{WHITE}{gray}{1}
 \definecolor{RED}{rgb}{1,0,0}
 \definecolor{GREEN}{rgb}{0,1,0}
 \definecolor{BLUE}{rgb}{0,0,1}
 \definecolor{CYAN}{cmyk}{1,0,0,0}
 \definecolor{MAGENTA}{cmyk}{0,1,0,0}
 \definecolor{YELLOW}{cmyk}{0,0,1,0}
 }

\usepackage{hyperref}
\hypersetup{
    colorlinks,%
    citecolor=blue,%
    filecolor=blue,%
    linkcolor=blue,%
    urlcolor=blue
}

\makeatother

\def\s{\sigma}

\def\p{\partial}

\newcommand{\be}{\begin{eqnarray}}
\newcommand{\ee}{\end{eqnarray}}

\begin{document}

\title{Exponential potential for an inflaton with nonminimal kinetic coupling and its supergravity embedding}

\author{Iannis Dalianis}
\email{dalianis@mail.ntua.gr}
\affiliation{ Physics Division, National Technical University of Athens, \\ 15780 Zografou Campus, Athens, Greece}  

\author{Fotis Farakos}
\email{fotisf@mail.muni.cz}
\affiliation{ Institute for Theoretical Physics, Masaryk University, \\  611 37 Brno, Czech Republic}


\begin{abstract}

In the light of the new observational results 
we discuss the status of the exponential potentials driving inflation. 
We depart form the minimal scenario and study an inflaton kinetically coupled to the Einstein tensor. We find that in this case the exponential potentials are well compatible with observations. Their predictions coincide with those of the chaotic type quadratic potential for an inflaton minimally coupled to gravity. We show that there exists a simple mapping between the two models. Moreover, a novel aspect of our model is that it features a natural exit from the inflationary phase even in the absence of a minimum.  We also turn to supergravity and motivate these sort of potentials and the non-minimal kinetic coupling as possible effective dilaton theories.

\end{abstract}


\maketitle
\section{Introduction}
Recently, the BICEP2 telescope has announced the observation of a B-mode polarization of the Cosmic Microwave Background (CMB) which for first time indicates a non-zero value for the  tensor-to-scalar ratio, $r$, at 7$\sigma$ C.L. \cite{Ade:2014xna}. This discovery, if confirmed by the future measurements, has striking cosmological implications.   The reported by BICEP2 values 
 together with the Planck data \cite{Ade:2013uln} provide a significant constraint on inflationary models. The value of $r$ is directly related to the scale of inflation and the type of the inflationary potentials. 
The large field models, see e.g. \cite{Baumann:2014nda,Lyth:1998xn,Langlois:2004de, Liddle:2000cg} for reviews, are well supported by the data. 
A particular example of large field models 
is the exponential potential which leads to power law inflation \cite{Lucchin:1984yf, Halliwell:1986ja,Burd:1988ss,Liddle:1988tb}. 
Potentials of this sort can arise in a number of microscopic theories of matter and interactions and in particular in stringy set ups, see Ref. \cite{Kitazawa:2014dya, Dudas:2010gi,Sagnotti:2013ica} for recent works, and Ref. \cite{Ashoorioon:2006wc, Ashoorioon:2008qr}.

However, cosmological inflation driven by exponential potentials and minimally coupled to gravity has been disfavoured by 
 the WMAP and Planck analysis, and, currently poorly fits the BICEP2 data. 

In this Letter, in the light of the new results, we re-draw our attention to the exponential potentials 
and in particular to an inflaton kinetically coupled to the Einstein tensor \cite{Amendola:1993uh, Gao:2010vr, Granda:2009fh, Saridakis:2010mf,Germani:2010gm,Germani:2010ux,Germani:2011ua,Germani:2014hqa}. 
We find that the predicted values of a kinetically coupled inflaton with exponential potential fit very well the observational data. Furthermore, the non-minimal kinetic coupling (or simply kinetic coupling) allows inflation to take place for a much wider range of values for the potential parameter $\lambda$, see Eq. (\ref{expV}). This feature has an important consequence: inflation can terminate contrary to the standard case where inflation never ends without the assistance of an additional mechanism. Once the kinetic coupling effects become negligible the potential is too steep to drive inflation and the inflaton's energy density rapidly redshifts. If the created radiation energy density becomes fast enough the dominant energy component in the post-inflationary universe then the standard thermal phase can be achieved.

A very interesting aspect of the theory for an inflaton with kinetic coupling and exponential potential is that the inflationary predictions coincide to leading order in the slow-roll parameter  with those of the chaotic type quadratic potentials. Although these two models appear to be unrelated we show that they are effectively described by the same dynamics during the inflationary phase. Actually, any potential for an inflaton with kinetic coupling can be transformed to a potential of another form for an inflaton minimally coupled to gravity. A similar observation concerning the Starobinsky theory was also made in the Ref. \cite{Kehagias:2013mya}. The exponential potential for an inflaton with kinetic coupling has observational signatures that distinguish it from the quadratic potentials: the post-inflationary evolution can be much different a fact that breaks the observational degeneracy.

The observational support along with the fact that exponential potentials emerge in generic stringy set-ups motivate us to incorporate this higher derivative theory into a supergravity framework. 
Let us mention in passing that there has been a renewed interest in embedding chaotic inflation in 
supergravity triggered by the new observational data  \cite{Ferrara:2014ima,Kallosh:2014qta,Hamaguchi:2014mza,Ellis:2014rxa,Kallosh:2014vja,Li:2014owa,
Ellis:2014gxa,Ferrara:2014fqa,Kallosh:2014xwa,Ferrara:2013rsa,Ferrara:2013kca, Burgess:2014tja, Farakos:2014gba}. 
Concerning our model, 
our result is that inflation driven by the kinetically coupled inflaton can be successfully described in terms of supergravity. We demonstrate that the kinetic coupling  does neither infer problematic instability issues nor the presence of intermediate strong energy scales for the graviton-inflaton system.

\section{Minimally coupled inflaton}
Let us assume an inflaton minimally coupled to gravity with an exponential potential
\begin{equation} \label{expV}
V=V_0 e^{-\lambda \phi/M_P}\,.
\end{equation}
The scale factor grows like $a(t)\propto t^{2/\lambda^2}$. In general the parameters $\lambda$ and $V_0$ should originate from some underlying theory that implies their values. 
When the exponential potentials are applied for implementing the early universe inflation it appears that they fit quite poorly the observational data. 
The slow-roll parameters 
\begin{equation} \label{slow}
\epsilon_V= \frac{M^2_P}{2} \left( \frac{V'}{V}\right)^2\,,\quad\quad \eta_V= M^2_P  \frac{V''}{V}
\end{equation}
yield the values $\epsilon_V=\lambda^2/2$ and $\eta_V=2\epsilon_V =\lambda^2$. An inflationary phase occurs only if $\lambda^2 <  2$.
This theory 
predicts the rather simple relation for the spectral index, $n_s$ and the tensor-to-scalar ratio, $r$,
\begin{equation} \label{mr}
 r=8(1-n_s)\,\quad \quad \text{\{\it{for minimally coupled inflaton}\}}
\end{equation}
where $1-n_s=6\epsilon_V-2\eta_V=\lambda^2$. The Planck collaboration \cite{Ade:2013uln} has excluded the exponential potential for the measured spectral index value $n_s=0.96$ signifies the presence of a strong signal of gravitational waves $r=0.32$, see Fig. 1 of \cite{Ade:2013uln}. This value is too large 
 even for the BICEP2 data \cite{Ade:2014xna}. The value $r=0.2^{+.07}_{-.05}$ gives $n_s=0.975^{-.009}_{+.006}$ 
 and when a dust reduction is taken into account, $r=0.16^{+.06}_{-.05}$, the $n_s$ value increases to $n_s=0.98^{-.007}_{+.006}$. These imply that the exponential potential for a minimally coupled inflaton cannot fit well the observaional data. They are actually too steep yielding a too large $\epsilon_V$. 
 
Apart from the indicated observational disproof there are also some theoretical difficulties. 
The coefficient $\lambda$ in the exponent has to be $\lambda= {\cal O}(0.1)$; for example the value $r=0.16$ requires $\lambda=0.14$. This means that the field value $\phi$ is suppressed effectively by a superplanckian mass $M_P/\lambda$ whereas, the case usually predicted by particle models, as e.g. in the string theory where the string and compactification scales appear, is that $\lambda \gtrsim 1$. A related issue is that a sufficient number of e-folds requires superplanckian excursions for the inflaton, $\Delta \phi=N\lambda M_P$. 
 In addition, it is well known that the exponential potential cannot account for a complete inflationary theory, since the slow-roll never ends and an additional mechanism is required to stop it. 

In the following we will show that these shortcomings can be addressed when the inflaton with an exponential potential has a kinetic coupling to the Einstein tensor.

\section{Inflaton with a kinetic coupling}

We shall consider a theory of a scalar field, that we identify it with the inflaton, kinetically coupled to gravity. The Lagrangian of this theory reads 
\begin{equation} \label {LA}
{\cal L}=\sqrt{-g} \left[\frac12 M^2_P R -\frac12\left(g^{\mu\nu} -\frac{G^{\mu\nu}}{\tilde{M}^2} \right)\partial_{\mu}\phi\partial_{\nu}\phi - V(\phi)\right]
\end{equation}
and $V=V_0\,e^{-\lambda\phi/M_P}$ for exponential potentials.
 In a homogeneous FLRW background the Friedmann equation and the equation of motion (EOM) for this kinetically coupled $\phi$ field are \cite{Germani:2010gm}
\begin{equation} \label{mod}
H^2=\frac{1}{3M^2_P}\left[\frac{\dot{\phi}^2}{2}\left(1+9\tilde{M}^{-2}H^2\right)+V(\phi) \right],
\end{equation}
\begin{equation} \label{mod2}
 \partial_t\left[a^3\dot{\phi}\left(1+3\tilde{M}^{-2}H^2 \right) \right]=-a^3 V_\phi\,.
\end{equation}
These equations yield the modified expressions for the slow-roll parameters
\begin{equation} \label{gefslow}
\epsilon=\frac{\epsilon_V}{1+3H^2\tilde{M}^{-2}}\,,\quad\quad \eta=\frac{\eta_V}{1+3H^2\tilde{M}^{-2}}\,,
\end{equation}
where the former is derived by the definition $\epsilon\equiv -\dot H/H^2$ and, the later by differentiating the approximated EOM form 
 where one finds that $\delta\equiv \ddot\phi/H\dot\phi=3\epsilon-\eta$. The $\epsilon_V$ and $\eta_V$ are the slow-roll parameters for the minimal case (\ref{slow}). When the new  scale $\tilde{M}$ is much smaller than the Hubble  scale, i.e. $H\tilde{M}^{-1} \gg 1$, we have the so called {\it high friction limit}. In this limit the Eq. (\ref{mod}) and (\ref{mod2}) take the form
\begin{equation} \label{fr}
H^2= \frac{1}{3M^2_P}\left[\frac32 \frac{\epsilon_V}{\epsilon}\dot{\phi}^2 +V(\phi) \right]
\end{equation}
\begin{equation} \label{mod2'}
\frac{\epsilon_V}{\epsilon}\left[3H\dot{\phi} +(3\epsilon-\eta)H\dot{\phi}-\frac{\dot{\epsilon}}{\epsilon}\dot{\phi} \right]=-V'
\end{equation}
where we used that $\ddot{\phi}=(3\epsilon-\eta)H\dot{\phi}$. During slow-roll inflation it is $\epsilon, \eta\ll 1$ and given that $\dot{\epsilon}=2H\epsilon^2$ the EOM Eq. (\ref{mod2'}) reads approximately
\begin{equation} \label{eom}
3H\dot{\phi} \simeq -\frac{\epsilon}{\epsilon_V} V'\,.
\end{equation}
The Eq. (\ref{eom}) implies, after straightforward calculations, that the Friedmann equation (\ref{fr}) reads
\begin{equation} \label{frsr}
H^2=\frac{1}{3M^2_P} \left[\frac{\epsilon V}{9} +V \right] \simeq \frac{V}{3M^2_P}
\end{equation}
and, indeed, during the slow-roll regime the potential dominates the energy density. The Eq. (\ref{frsr}) and (\ref{eom}) are the master equations that describe the inflationary dynamics in the case of kinetic coupling.

In the high friction limit the spectral tilt of the scalar power spectrum has, in turn, a modified dependence on the slow-roll parameters \cite{Germani:2010ux, Germani:2011ua}
\begin{equation} \label{ns}
\left. n_s-1 \equiv \frac{d \ln {\cal P}_R}{d \ln k}\right|_{c_sk=aH} \simeq -8 {\epsilon} + 2{\eta}
\end{equation}
due to the sensitivity of the ${\cal P_R}$ to the modulated by the first slow-roll parameter sound speed of the perturbations \cite{Germani:2011ua}. The high friction limit, albeit, leaves the ${\cal P_R}$ amplitude and, in first order, the tensor amplitude $P_t$ unchanged. Therefore the tensor-to-scalar power ratio in terms of the slow-roll parameter $\epsilon$ has approximately the standard form. 
 Summing up, in the case of kinetic coupling and for the exponential potential the  $n_s$ and the $r$  are related to leading order to $\epsilon$ as follows:  
\begin{equation} \label{pred}
\left. \begin{matrix} 
r\simeq16\epsilon & \\
1-n_s \simeq 4\epsilon &\\
r \simeq 4(1-n_s) & 
\end{matrix} \right.
 \quad \text{\{\it{for an inflaton with kinetic coupling}\}}
\end{equation}
Comparing with the expression (\ref{mr}) for the minimal case we see that, here, the predicted value for the tensor-to-scalar power ratio is {\it lower} for a given $n_s$. We illustrate the different behaviour of the $r=r(n_s)$ for the minimal and the kinetic coupling cases in the Fig. 1 where the BICEP2 and the Planck (for running spectral index) contours are in the background. The exponential potential in the minimal case predicts zero running and is excluded by the Planck data. 

The expression for the number of e-folds before the end of inflation, $N_*$,  at which the pivot scale $k_*$ exits the Hubble radius, is also modified. It reads
\begin{equation} \nonumber
N_*\equiv\int^{t_*}_{t_e} Hdt =\frac{1}{M^2_P} \int^{\phi_*}_{\phi_e}(1+3H^2\tilde{M}^{-2}) \frac{V}{V,_\phi}d\phi
\end{equation}
\begin{equation} \label{N*}
=\frac{\phi_e-\phi_*}{\lambda M_P}\,+\,\frac{V_0}{\lambda^2 M^2_P \tilde{M}^2} \left(e^{-\lambda \phi_*/M_P}-e^{-\lambda \phi_e/M_P}\right)\,.
\end{equation}

\begin{figure} \label{X}
\centering
\includegraphics [scale=0.6, angle=0]{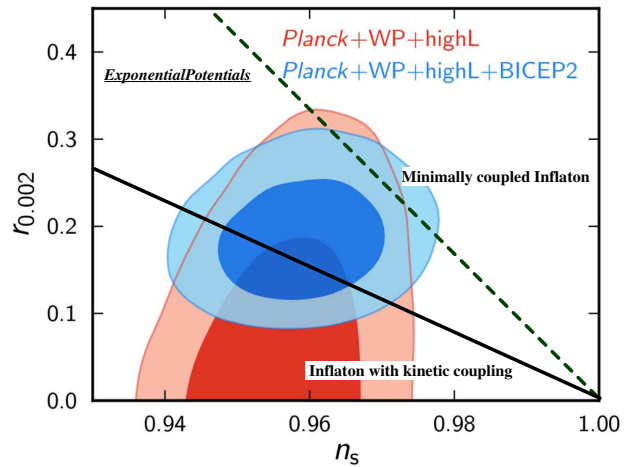}  
\caption{Theoretical predictions for $n_s$ and $r$ for an inflaton with standard exponential potential (dashed line) and an inflaton kinetically coupled to gravity (solid line) compared to the BICEP2 data \cite{Ade:2014xna}. The red contours constrain models with tensors and, additionally, with running of the spectral index 
$dn_s/d\ln k\simeq -0.022$. Both lines should be compared with the BICEP2 {\it blue contour}, since both models predict either strictly zero (dashed line) or close to zero, ${\cal O}(10^{-3})$, (solid line) running.}
\end{figure}

\subsection{Inflation terminates}
Inflation takes place for $\epsilon<1$. From the expressions (\ref{gefslow}) in the high friction limit we take  that $\epsilon\simeq \epsilon_V/(3H^2\tilde{M}^{-2})=\lambda^2/(6H^2\tilde{M}^{-2})$. Hence, a first comment is that inflation can occur for much larger values for the parameter $\lambda$ than in the standard case, i.e. $\lambda \gg {\cal O}(0.1)$.  Plugging in the Hubble scale parameter during inflation, $3H^2 \simeq V(\phi)/M^2_P$, we find that inflation is realized only for sufficient large values for the potential
\begin{equation}
V(\phi)>\frac12{(\lambda \tilde{M}M_P)^2}\,.
\end{equation}
The above condition reads in terms of the inflaton field
\begin{equation} \label{phie}
\phi < \frac{M_P}{\lambda} \ln \left(\frac{2V_0}{\lambda^2 \tilde{M}^2 M^2_P } \right) \,.
\end{equation}
Thereby, in the case of the kinetic coupling an inflation with an exponential potential {\it does have an end} contrary to the minimal case. 
Indeed, we see from the expressions for the slow-roll parameters (\ref{gefslow}) that when inflation ends, $\epsilon=1$, the Hubble scale has a value $H_e$
\begin{equation}
H^2_e=(\lambda^2-2)\frac{\tilde{M}^2}{6}\,.
\end{equation}
Unless $\lambda> \sqrt{2}$ inflation is endless. Actually, in the high friction limit it is $H^2_e\gg \tilde{M}^2$ which implies that 
$\lambda^2 \gg 1$. 
When the friction effects become negligible, $H\lesssim \tilde{M}$, the $\phi$-inflaton field evolves in the standard way. It has been shown \cite{Liddle:1988tb} that for $\lambda^2>6$ the system tends to a solution where the scale factor has a power-law behaviour with exponent $1/3$, $a\propto t^{1/3}$. This corresponds to stiff matter behaviour, $p=\rho$, with energy density scaling like $\rho \propto a^{-6}$. Therefore given that the field $\phi$, after the end of inflation,  has transferred sufficient part of its energy density into entropy it will remain a fast redshifting and subdominant component of the universe energy density. In other words the absence of a minimum, which appears to be a negative feature of the exponential potential, may {\it not} be problematic in kinetically coupled to gravity inflaton scenarios.
 Nevertheless, we do not claim that this model should not receive further completion. An additional mechanism may exist that generates a minimum, however, 
there is no need here the position of this minimum to be necessarily linked with the end of inflation.

\subsection{The predicted values for the inflationary observables}
{\it The tensor-to-scalar ratio}. 
The scalar spectral index value measured by Planck is $n_s=0.96$. Inflation driven by an inflaton kinetically coupled to gravity and with an exponential potential predicts (\ref{pred})
\be
r=0.16
\ee
when we plug in  $n_s=0.96$. This value is in well agreement with the BICEP2 data. \\
\\{\it The number of e-folds}.
The potential energy during inflation is 
$V(\phi_*)=V_0\,e^{-\lambda \phi_*/M_P}=3.3\times 10^{-8}r_*M_P^4 $
which gives the following value for the exponent:
\begin{equation} \label{phi*}
\lambda \phi_*=\left[ 19 -\ln \left(\frac{r_*}{0.16}\right) +\ln \left(\frac{V_0}{M^4_P}\right) \right] M_P\,.
\end{equation}
For $r\sim 0.16$ and $V_0=M^4_P$ 
 the exponent takes the value $\lambda \phi_* \simeq 19 M_P$. 
Taking into account that in the high friction limit it is $\lambda \gg 1$, the relations (\ref{phie}), (\ref{phi*}) and the value of the $\Delta \phi=\phi_e-\phi_*$ from Eq. (\ref{dphi}) we find that the number of e-folds are approximately given by the second term of (\ref{N*}) i.e.
\begin{equation} \label{N2}
N_* \simeq \frac{V(\phi_*)-V(\phi_e)}{\lambda^2 M^2_P \tilde{M}^2} 
\end{equation}
During slow-roll inflation in the limit $3H^2 \gg \tilde{M}^2$ we can write $H^2\simeq [\dot{\phi}^2(1+3\epsilon_V/\epsilon)/2+V]/3M^2_P$ $\simeq V/3M^2_P$. Thereby, from the first slow-roll parameter value at the pivot scale $\epsilon_*=(\lambda \tilde M)^2/6H^2_*\simeq (\lambda \tilde{M}M_P)^2/2V(\phi_*)$ and the end of inflation $\epsilon_e\simeq (\lambda \tilde{M}M_P)^2/2V(\phi_e)=1$ we find that
\begin{equation} \label{VN}
V(\phi_*) \simeq V(\phi_e)(2N_*+1)
\end{equation}
and 
\begin{equation}
\epsilon_* \simeq \frac{V(\phi_e)}{V(\phi_*)} = \frac{1}{2N_*+1}\,.
\end{equation}
Hence, using the tensor-to-scalar ratio $r_*=16\epsilon_*$ we take the following relations 
at the pivot scale:
\begin{equation} \label{rN}
r_* \simeq 8 \frac{(\lambda \tilde{M} M_P)^2}{V(\phi_*)} \simeq \frac{16}{2N_*+1}\,
\end{equation}
and
\begin{equation}
1-n_s=\frac{4}{2N_*+1}\,.
\end{equation}
If we ask $N_*\leq 60$ we take that $r_* \geq 0.135$. The relation (\ref{rN}) accounts for a prediction of the exponential type potentials with an  inflaton kinetically coupled to gravity. For $r_*=0.16$ we take $N_*\simeq 50$.
\\
\\
{\it The parameter $\lambda$ and the mass scale $\tilde{M}$}.
We can recast the expression (\ref{rN}) in such a way that the product $\lambda\tilde{M}$ is given in terms of measured quantities $H_*$,  $r_*$
\begin{equation} \label{lM}
\lambda \tilde{M} = \sqrt{\frac{3}{8}} H_* \,r^{1/2}_*\, .
\end{equation}
Also, given the measured amplitude for the scalar power spectrum, ${\cal P}_R$, the Hubble parameter reads $H_*=1.05 \times 10^{-4} r^{1/2}_* M_P$ and the (\ref{lM}) is rewritten as  
 $\lambda \tilde{M} = 6.4 \times 10^{-5} r_* M_P.$ 
 Therefore the free mass scale parameter $\tilde{M}$ is directly related to the parameter $\lambda$ of the inflationary potential. For example if $\lambda \sim 10$ the suppression scale of the non-minimal derivative coupling is $\tilde{M}\sim 10^{-6} M_P$.
\\
\\
{\it The running of the spectral index}. 
In the high friction limit we find that 
\begin{align}
\begin{split}
\frac{dn_s}{d\ln k} &= 
-4\frac{d\epsilon}{d\ln k}=-2 \left(\frac{\lambda^2 \tilde{M}^2 M^2_P}{V(\phi_*)}\right)^2\\ &
\simeq -\frac{8}{(2N_*+1)^2}\,.
\end{split}
\end{align}
Although there {\it is} a running of the spectral index here, contrary to the minimal case, the $n_s$ running is not that significant so as to reconcile the different data according to the BICEP2 suggestion \cite{Ade:2014xna}. 
\\
\\
{\it The variation of the field value during inflation}. 
According to (\ref{phie}) the value of the field at the end of inflation is $\phi_e=(M_P/\lambda)\ln(2V_0/\lambda^2\tilde{M}^2M^2_P)$. Plugging in the value $\lambda \tilde{M} = 6.4 \times 10^{-5} r_* M_P$  we find
\begin{equation}
\phi_e \simeq \frac{M_P}{\lambda} \left[23.6-2\ln \left(\frac{r_*}{0.16} \right) +\ln\left(\frac{V_0}{M^4_P} \right) \right]\,.
\end{equation}
Hence
\begin{equation} \label{dphi}
\Delta \phi =\phi_e-\phi_* \simeq   \left[4.6- \ln \left(\frac{r_*}{0.16} \right)\right]\frac{M_P}{\lambda}
\end{equation}
where $\phi_*$ 
is given by the expression (\ref{phi*}).
Although the variation of the $\phi$ field is possibly well subplanckian we emphasize that the $\phi$ field is {\it not} canonical.  Actually, only during an exact de-Sitter phase one can recast the kinetic term for the scalar field (\ref{LA}) into a canonical form, where  $\tilde{\phi}= \sqrt{3}H\tilde{M}^{-1}\phi$ is the canonically normalized field and $H=$ constant. Inflation is a quasi de-Sitter phase where $\dot{H}=-\epsilon H^2\ll 1$. Hence, a crude estimate $\Delta\tilde{\phi}\sim \sqrt{3}H\tilde{M}^{-1}\Delta \phi \sim 4.6\sqrt{8}\,r^{-1/2}_* (H/H_*)M_P$, where the expression (\ref{lM}) has been used, indicates that the canonical field variation is superplanckian
and the Lyth bound \cite{Lyth:1996im} is not violated, see also \cite{Kehagias:2014wza, Dvali:2014ssa, Chialva:2014rla}.  
\\
\\
We note that, despite the new nonliner interaction of gravity to the inflaton field, {\it no non-Gaussian} fluctuations larger than those in general relativity are produced. It has been actually shown \cite{Germani:2011ua} that in the single field case the non-Gaussianities of curvature perturbations are suppressed by slow-roll as in \cite{Maldacena:2002vr}.
\\
\\
Let us now briefly comment on the above results. The condition $\lambda \gg 1$, implied by the observational data, is welcome for several reasons. Firstly, we have a subplanckian suppression scale $M_P/\lambda$, which in particular models can be identified e.g. with the string scale or the compactification scale. Secondly, the (non-canonical) field $\phi$ experiences subplanckian excursions which may be prerequisite in some more complete models that  incorporate this effective theory. Moreover, in the post-inflationary phase when $H\leq \tilde{M}$ 
 the energy density of the inflaton redshifts faster than radiation without spoiling the hot Big-Bang scenario (thermal phase, BBN for $\lambda^2 \gtrsim20$ \cite{Ferreira:1997hj,Peebles:1998qn}). Here, we have assumed that a sufficient reheating has taken place.
 On the other hand, if $\lambda^2< 2$  
  inflation never ends. 

We mention that the presence of a new intermediate scale, $\tilde{M}$, introduces new non-renormalizable interactions. However, in a single field inflation with kinetic coupling the strong coupling scale of the inflaton-graviton system can be identified with the Planck scale \cite{Germani:2011ua, Germani:2014hqa}. This fact implies that the standard calculations for the inflationary phase can be safely carried out and the results presented in this section are reliable.

\section{Effective Supergravity Description}

In this part, we want to embed our results into supergravity. 
Exponential potentials are of particular interest in supergravity since they 
are connected to the superstring dilaton.  
In fact previous work has pointed out some interesting aspects of the model. 
First, the non-minimal derivative coupling has only been constructed in the framework of the new-minimal supergravity \cite{Sohnius:1981tp,Cecotti:1987qe,Ferrara:1988pd,Ferrara:1988qxa}, 
a supergravity which is believed to be related to the heterotic string (for a discussion see for example \cite{Cecotti:1987mr}). 
In particular, the specific coupling can be found among the couplings of the dilaton in the effective 
heterotic string action \cite{Gross:1986mw}. 
Moreover, 
as has been pointed out by earlier work, 
it is not possible to explicitly introduce a superpotential for this field in a new-minimal supergravity framework, 
since it is forbidden by the R-symmetry \cite{Farakos:2012je,Farakos:2013fne}. 
Thus, as has been shown recently, 
the only consistent self-coupling of our superfield is via the gauge-kinematic function, 
and eventually this generates a potential \cite{Dalianis:2014sqa}. 
Interestingly enough, this is how the dilaton superfield is expected to couple to the gauge superfields. 

It is then evident that the model studied in the previous sections, which fits well the observational data, deserves a thorough investigation 
in the framework of new-minimal supergravity. 
Inflationary models in supergravity with chiral superfields and higher derivative couplings 
can be also found in \cite{Koehn:2012np,Gwyn:2014wna}.

The  new-minimal supergravity  \cite{Sohnius:1981tp} is the supersymmetric theory of the gravitational multiplet
\begin{equation}
e^a_m\, ,~~\psi_m^\alpha \,  , ~~A_m\, , ~~ B_{mn}\ .
\end{equation}
The first two fields are the vierbein and its superpartner the gravitino, 
a spin-$\frac{3}{2}$ Rarita-Schwinger field. 
The last two fields are auxiliaries. 
The real auxiliary vector $A_m$ gauges the $U(1)_\text{R}$ chiral symmetry. 
The auxiliary $B_{mn}$ is a real two-form appearing only through its dual field strength $H_m$, 
which satisfies  
\begin{equation}
\label{DH}
\hat {\cal D}^a H_a =0  
\end{equation}
for the supercovariant derivative $\hat {\cal D}^a$.

A special feature of new-minimal supergravity is that is contains a gauged R-symmetry, $U(1)_\text{R}$. 
This R-symmetry places restrictions on the supersymmetric theories one can write down. 
To illustrate this let us take a short detour through the global theory. 
The global R-symmetry acts on the theta parameters as
\be
\label{Rtheta}
\text{R} \ \theta^\alpha = e^{-i \phi} \theta^\alpha 
\rightarrow 
\text{R} \ d\theta^\alpha = e^{-i \phi} d\theta^\alpha
\ee
and on a chiral superfield as
\be
\label{RPhi}
\text{R} \  \Phi(x,\theta)  = e^{2in \phi} \Phi(x, e^{-i \phi} \theta) 
\ee
where $n$ is the R-charge. 
A renormalizable supersymmetric theory reads
\be
{\cal L} = \int d^4 \theta \Phi \bar \Phi + \int d^2 \theta W(\Phi) +c.c.
\ee
for
\be
\label{RSP}
W(\Phi) = f \Phi + m \Phi^2 + \lambda \Phi^3 . 
\ee
It is then easy to realize from (\ref{Rtheta}) and (\ref{RPhi})  that in a theory that respects the R-symmetry, 
only one of the terms in the superpotential (\ref{RSP}) would be allowed. 
For example 
\be
W(\Phi) = m \Phi^2 \rightarrow n=1 . 
\ee
In general the superpotential carries R-charge $n=2$. 
In new-minimal supergravity this R-symmetry becomes local and is gauged by the auxiliary field $A_m$, 
which transforms as
\be
\delta_\text{R} A_m= \p_m \gamma .  
\ee

We are now ready to incorporate the aforementioned models of inflation into supergravity. 
The inflaton field will be accommodated into a chiral superfield $\Phi$. 
As has been found in earlier work \cite{Farakos:2012je}, 
in order to consistently couple the chiral superfield to curvature via a 
non-minimal derivative coupling, 
it has to carry a vanishing R-charge, 
but from the previous discussion this tells us that  a superpotential for this superfield is 
forbidden, since a superpotential has to bear R-charge $n=2$. 
On the other hand, if there exist other chiral superfields with a non-vanishing R-charge 
coupled to supergravity, 
this will again lead to ghost instability due to the fact that the auxiliary field $A_m$ will seize to be 
a Lagrange multiplier (giving the equations of motion for $H_a$)  and will have quadratic terms. 
Thus one may not introduce any R-charged chiral superfield in this model, 
and thus no superpotential. 
It has been found that there exists an indirect way to introduce a potential for 
our superfield via a gauge kinematic function, 
thus extending D-term inflation \cite{Binetruy:1996xj,Binetruy:2004hh} 
to higher derivative D-term inflation  \cite{Dalianis:2014sqa}.

In order for our Lagrangian to be manifestly supersymmetric, 
we derive it in  a superspace framework \cite{Ferrara:1988qxa}. 
Our model  in superspace is  
\be
\label{L1}
\begin{split}
{\cal L} =& -  2 M_P^2 \int d^4 \theta E V_\text{R} +  \int d^4  \theta \, E \,  K( \bar \Phi, \Phi) 
\\
& + i M^{-2}_* \int d^4 \theta \, E \, [ \bar \Phi E^a \nabla_a \Phi ] + c.c.  
\\
& +\frac{1}{4} \int d^2 \theta {\cal E} f(\Phi) W^2 (V) + c.c. 
\\
&+ 2 \xi \int d^4 \theta E V\,,
\end{split}
\ee
where $V_\text{R}$ is the gauge superfield of the R-symmetry and it also carries the Ricci scalar on its 
highest component, $V$ is a $U(1)$ gauge superfield and 
\be
W_\alpha(V) = -\frac{1}{4} \bar \nabla^2 \nabla_\alpha V
\ee 
is the standard field strength chiral superfield in new-minimal supergravity.

The K\"ahler potential $K(\Phi, \bar \Phi)$ in (\ref{L1}) will be considered to be canonical, nevertheless other forms are possible. 
In fact during inflation the form of the K\"ahler potential is not relevant. 
We write $z$ and $F_{mn}$ respectively:
\be
 z=\Phi|=\phi + i\beta
\ee 
the dynamical scalar component of the chiral superfield $\Phi$ and, $F_{mn}$ the 
field strength of the dynamical $U(1)$ gauge field 
\be
C_m = \frac{1}{4} \bar \s^{\dot \alpha \alpha}_{m} [ \nabla_\alpha , \bar \nabla_{\dot \alpha} ] V | . 
\ee 
Then the component form of the superspace Lagrangian (\ref{L1}), 
after integrating out the auxiliary sector  reads  \cite{Dalianis:2014sqa}
\begin{equation}
\label{total-onshell}
\begin{split}
e^{-1}{\cal{L}}=& \frac{M^2_P}{2} {{\cal{R}}}  + z\Box \bar z 
+\tilde{M}^{-2}
\, G^{ab}\,\partial_a 
\bar z\, \partial_b z\ -\frac{1}{2} \frac{\xi^2}{\text{Re}f(z)} 
\\
& -\frac{1}{4}  \text{Re}f(z) F^{mn} F_{mn} 
+ \frac{1}{4} \text{Im}f(z) F^{mn} \ ^*F_{mn}\,,
\end{split}
\end{equation}
where $\xi$ is the Fayet-Iliopoulos (FI) parameter of mass dimension two. Note that this Lagrangian (\ref{total-onshell}) does {\it not} contain ghost states or instabilities.

The theory we have discussed in the previous sections is reproduced by setting
\begin{equation} \label{fexp}
f(z)= \frac{\xi^2}{2V_0} e^{\lambda z/M_P}
\end{equation}
then we directly get an exponential type potential
\begin{equation} \label{exp}
{\cal V} =\frac12 \frac{\xi^2}{\text{Re}f(z)}=V_0 \frac{1}{\cos(\lambda \beta/M_P)} e^{-\lambda \phi/M_P}\,.
\end{equation}
The form of the function $1/\cos x$ implies that the $\beta$-dependent part of the potential will be stabilized 
to values  $\langle{1/\cos}(\lambda \beta/ M_P) \rangle = 1 $ and the $\phi$ will be the inflating field.
\\
{\it The mass of the imaginary field}.
It is important the effective theory to be a single field one. Otherwise new energy scales in the intermediate appear \cite{Germani:2011ua}. We want the imaginary field to be  integrated out. Its mass in the $\beta$-direction minimum is
\begin{equation}
\left. m^2_\beta \equiv \frac{\partial^2 V}{\partial \beta^2}\right|_{\beta=0} = V_0 \frac{\lambda^2}{M^2_P} e^{-\lambda\phi/M_P} \simeq 3 \lambda^2 H^2\,.
\end{equation}
Therefore if $\lambda \gg 1$ then the non-inflating $\beta$ field is decoupled and the effective theory is a single field one. In a single field theory with non-minimal derivative coupling the strong coupling scale of the graviton-inflaton system is determined by the graviton only interactions and can be identified approximately with the Planck mass \cite{Germani:2011ua, Germani:2014hqa}. Hence, the single field results of the section III remain valid here as well where the inflaton is the real part of the complex scalar $z=\phi+i\beta$.

In addition we can assume that the $C_m$ field where, $F_{mn}=\partial_mC_n-\partial_nC_m$, is coupled to charged matter in $U(1)$ gauge invariant manner. Then, rewriting the Lagrangian in terms of the cannonically normalized field the physical coupling will scale as $\propto 1/\text{Ref(z)}$. Assuming that at the end of inflation the gauge coupling is less than one then there is no any strong coupling problem during inflation according to the form of the potential (\ref{exp}).

Let us finally note that the FI term in (\ref{L1}) is not the only way to introduce a potential. 
In fact since the superfield $\Phi$ appears in the gauge kinematic function, 
one may instead of the $U(1)$ sector of  (\ref{L1}), introduce a hidden strongly coupled sector 
\be
{\cal L}_{sc} = \frac{1}{4} \int d^2 \theta {\cal E}  \frac{1}{g^2} e^{- \tilde{\lambda} \Phi/M_P} W^2 (V') + c.c. 
\ee
This sector, if we allow for  a non-vanishing scalar condensate of the gauge fermions \cite{Ferrara:1982qs,Nilles:1990zd,Kim:1991mv} 
\be 
\langle \lambda'^\alpha \lambda'_\alpha \rangle =\Lambda^3\,,
\ee 
will lead to a non-perturbative superpotential
\be \label{wnp}
W_{NP} =  \frac{\Lambda^3}{4g^2} e^{- \tilde{\lambda} \Phi/M_P}\,.
\ee
The (\ref{wnp}) will introduce a potential of the form
\be
{\cal V} = \frac{\Lambda^6 \tilde{\lambda}^2}{16 g^4 M_P^2} 
e^{-{\tilde{\lambda}(z + \bar z)}/{M_P}}\,.
\ee
This potential may have interesting cosmological implications. Here we mention that such kind of non-perturbative effects may be used for stabilizing the inflaton runaway direction \cite{King:1998uv}. 
However, as we emphasized in the text, the stabilization may not be necessary and it is disconnected to the end of inflation.

\section{Comparison to the quadratic chaotic models}
\subsection{Similarities}
The predictions of the inflationary model characterized by an exponential potential for an inflaton with a kinetic coupling are astonishingly similar to the predictions of the quadratic chaotic potential $V_{\varphi^2}(\phi)=m^2\varphi^2/2$ where $\varphi$ minimally coupled to gravity. The spectral index, the tensor-to-scalar ratio and the running of the spectral index read in the case of a $\varphi^2$ potential 
\begin{align} \label{qua}
\begin{split}
& 1-n_{s,\varphi^2}\simeq \frac{4}{2N_*+1} \,,\\
& r_{*,\varphi^2}\simeq \frac{16}{2N_*+1} \,,\\
& \frac{dn_{s,\varphi^2}}{d\ln k} \simeq -\frac{8}{(2N_*+1)^2}
\end{split}
\end{align}
which are exactly the same with those of the section III. 
This coincidence of the predictions can be also seen from the form of the $\epsilon_{\varphi^2}$ and $\eta_{\varphi^2}$ slow-roll parameters
\begin{equation} \label{ep}
\epsilon_{\varphi^2}\equiv\frac{M^2_P}{2}\left(\frac{V'_{\varphi^2}}{V_{\varphi^2}}\right)^2=\frac{2 M^2_P}{\varphi^2} =\frac{ m^2 M^2_P}{ V_{\varphi^2}}\,,
\end{equation}
\begin{equation}\label{ee}
\eta_{\varphi^2}\equiv M^2_P \frac{V''_{\varphi^2}}{V_{\varphi^2}} = \epsilon_{\varphi^2}\,.
\end{equation}
The (\ref{ep}) and (\ref{ee}) are of the same form with the $\epsilon$ and $\eta$ slow-roll parameters for exponential potentials for an inflaton with kinetic coupling. Indeed it is 
\begin{equation}
\epsilon \simeq \frac{\lambda^2 \tilde{M}^2 M^2_P}{2V(\phi)} \,\longleftrightarrow\, \epsilon_{\varphi^2}
\end{equation}
for 
\begin{equation}
\lambda^2 \tilde{M}^2=2m^2\,.
\end{equation}
The number of e-folds $N_*$ have also a similar dependence. Hence, the relations (\ref{pred}), between the $r$ and $n_s$ of the section III, hold for the quadratic  potential as well. We add that both theories do not generate important non-gaussianities.
These facts imply that there is an underlying relation between these two models that we are investigating in the following.

\subsection{Correspondence between the dynamics of an inflaton with kinetic coupling and a minimally coupled inflaton}

When the inflaton field $\phi$ has non-minimal derivative coupling to the Einstein tensor during slow-roll regime and in the high friction limit $H^2\gg \tilde{M}^2$ its dynamics are governed by 
the system of the 
equations (\ref{frsr}) and (\ref{eom}):
\begin{align} \label{syst}
H^2\simeq \frac{V(\phi)}{3M^2_P}\,,\quad\quad 3H\dot{\phi} \simeq - \frac{\epsilon}{\epsilon_V}V'(\phi)\,.
\end{align}
According to the previous subsection there is a clear hint of a correspondence between the non-minimally coupled inflaton with exponential potential and the minimally coupled inflaton with quadratic potential. There should be a {\itshape generic} transformation of the form
\begin{align}
\varphi=g(\phi), \quad\quad V_m(\varphi)=V[g^{-1}(\varphi)]
\end{align}
such that the above system of equations (\ref{syst}) is recast into
\begin{equation} \label{msyst}
H^2\simeq \frac{V_m(\varphi)}{3M^2_P},\quad\quad 3H\dot{\varphi} \simeq -V'_m(\varphi)\,,
\end{equation}
where $V_m(\varphi)$ a potential for the field $\varphi$ minimally coupled to gravity. After straightforward calculations the EOM of (\ref{msyst}) is written in terms of the $\phi$ field as 
\begin{align}
3H\dot{\phi} \simeq -\frac{V'(\phi)}{[g'(\phi)]^2}\,,
\end{align}
where the prime denotes derivative with respect to the argument field.  This equation is equivalent to the EOM of the system (\ref{syst}) if $[g'(\phi)]^2=\epsilon_V/\epsilon$ or 
\begin{align}
g'(\phi)= \frac{V^{1/2}}{M_P\tilde{M}}\,.
\end{align}
Therefore the new field $\varphi$ reads in terms of the field $\phi$
\begin{equation} \label{trans}
\varphi=\int\frac{V^{1/2}}{M_P\tilde{M}}\,d\phi\,.
\end{equation}
This formula can be also found in \cite{Germani:2011mx}. Substituting $V=V_0e^{-\lambda \phi/M_P}$ we take
\begin{align} \label{fff}
\varphi= -\frac{2V^{1/2}_0}{\lambda \tilde{M}} e^{-\lambda \phi/2M_P}
\end{align}
and, as well, the inverse function $g^{-1}(\varphi)=\phi=-(2M_P/\lambda)$ $\times$ $\ln(-\lambda \tilde{M}\varphi/2V^{1/2}_0)$. It follows that  the potential $V_m$ for the minimally coupled $\varphi$ field reads
\begin{align}
V_m(\varphi)=V[g^{-1}(\varphi)]=\frac12\frac{\lambda^2\tilde{M}^2}{2}\varphi^2\,.
\end{align}
Hence, the exponential potential is mapped to the quadratic potential:
\begin{align}
V_m(\varphi)\equiv V_{\varphi^2} (\varphi) = \frac12 m^2 \varphi^2\,,
\end{align}
where $m^2=\lambda^2\tilde{M}^2/2$ with the $\varphi$ field governed by 
 the standard slow-roll regime system of equations:
\begin{align} \label{ssyst}
H^2\simeq \frac{V_{\varphi^2}(\varphi)}{3M^2_P}\,,\quad\quad 3H\dot{\varphi} \simeq - V'_{\varphi^2}(\varphi)\,.
\end{align}
We also see that, e.g., for 50 e-fold of inflation for the minimally coupled field $\varphi$ \cite{Langlois:2004de} one also takes 50 e-folds of inflation for the non-minmally coupled field $\phi$, see formuli (\ref{dphi}) and (\ref{fff}). The excursion for the field $\varphi$ is superplanckian.

Summing up, the system of equation (\ref{syst}) for an inflaton with exponential potential and kinetic coupling is recast into the system (\ref{ssyst}) after the transformation (\ref{trans}). 
 The system (\ref{ssyst}) describes the slow-roll inflation driven by the minimally coupled field $\varphi$ characterized by "chaotic-type" quadratic potential. Therefore, the predictions of these two theories naturally coincide and any differences are expected to be at the level of the slow-roll parameters $\epsilon$ and $\eta$.

\subsection{Differences and potential observational signatures}
The essential point is the search for observational signatures that break the degeneracy between the two models. The period between the end of inflation and the initiation of the radiation dominated era plays here a crucial role since each model predicts a different cosmic post-inflationary evolution.
The exponential inflationary potential fits well the data only if it is too steep for conventional inflation to occur. It is the non-minimal derivative coupling acting in the very high energies that enables the implementation of the slow-roll regime. At the end of inflation the Hubble parameter has the value $H_e \sim \lambda \tilde{M}$. Given that $\lambda>\sqrt{6}$ the equation of state will gradually increase from the value $w=-1/3$ at the end of inflation to the stiff matter value $w=1$ due to the steepness of the potential and the absence of a minimum. This is the decisive difference between the quadratic chaotic models and the one presented here. 

{\it Entropy production.} In order the standard hot Big Bang scenario to be realized some of the inflaton energy needs to be converted to radiation. This may happen via three different ways. The first is the gravitational particle production \cite{Ford:1986sy} where the entropy is produced due to the variation of the scale factor $a(t)$ with time, i.e. due to the time varying gravitational field. The density of particles produced at the end of inflation is found to be
\begin{equation}
\rho_R \sim 0.01\, g_p \,H^4_e
\end{equation}
where $g_p$ is the number of different particle species created from the vacuum. The common expectation is that $g_p \sim 10-100$; the $H_e$ reads in terms of observable quantities $H_e \,(2N_*+1)^{1/2} \simeq H_*\simeq 10^{-4}r^{1/2}_*M_P$ where the Eq. (\ref{VN}) was used. Hence the radiation is a small fraction of the total energy, $\rho_R/\rho_\phi \sim 10^{-10}g_p\, r_*/6N_*$. 
 Once the kinetic regime ($w=1$) 
 commences the radiation energy density will increase relatively to the dominant $\phi$ field energy density as 
\begin{equation} 
\frac {\rho_R}{\rho_\phi} \propto a^2(t)\,.
\end{equation}
The radiation domination era will take over the stiff matter era before the BBN, however,  there is a stronger constraint coming from the backreaction of gravity waves, as we mention in the following.

A second way that entropy production can be accomplished is  via instant preheating \cite{Felder:1999pv}. This requires a coupling of the form 
\begin{equation}
\delta {\cal L} = h^2 \phi^2 \chi^2
\end{equation}
and further Yukawa couplings of the $\chi$ particles to other matter fields. After the production of the $\chi$ particles their effective mass $h\phi$ grows together the $\phi$ field value. The energy density of the $\chi$ particles and of the products of their decay can soon become important.  Hence, this coupling may result in a much faster and sufficient reheating of the universe. 

Finally, a third way to implement the post-inflationary thermal phase is to proceed to a completion of the theory via terms that can generate a minimum to the potential. When $\phi\sim \phi_{min}$ the $\phi$ field oscillates about the minimum and eventually decays (assuming no overshooting). A study of the decay process when non-minimal derivative couplings are present has been performed in \cite{Koutsoumbas:2013boa}.  Nevertheless, the position of the minimum can be at field values $\phi_{min} \gg \phi_e$ and the above discussion remains relevant for this case too. 

{\it Gravitational waves.} The existence of a post-inflationary phase stiffer than radiation influences the relative amplitude and the tilt of the stochastic gravitational waves \cite{Giovannini:1998bp}. The energy density of relic gravity waves scales like $\rho_g \propto a^{-4}$ when the background energy density is characterized by $w>1/3$ and, mimicks the background scaling in the opposite case \cite{Sahni:2001qp}. This implies that during the stiff matter phase the energy density in the gravity waves will increase relative to the background, a fact that does not permit the scalar field stiff matter period to extend too long. In addition a prolonged stiff matter regime affects the gravity wave spectrum giving it considerable power on short wavelength scales. As a result of such a case the gravity waves will have a blue spectrum.  A sufficient entropy production, on the other hand, shortens the kinetic phase domination and hence decreases the backreaction impact and the scale-dependent tilt of the gravity waves. 

{\it Gauge fields.} The inflaton interaction with the gauge fields, see the Lagrangian (\ref{total-onshell}), is a particular feature of the supergravity inflationary models with kinetic coupling. The time dependence of the gauge kinetic function breaks the conformal invariance of the gauge field sector and leads to amplification of the quantum fluctuations of the gauge bosons during the nearly de-Sitter phase. Furthermore, the coupling of the inflaton to a gauge field also contributes to the reheating of the universe \cite{Ferrara:2011dz}.

In summary, the exponential potential for an inflaton with kinetic coupling predicts the same values for the cosmological perturbation parameters. However, the post-inflationary phase may be much different and modify the predictions. The end of inflation is followed by a stiff matter kinetic regime and, hence, the radiation domination phase may start much later.  This fact alters the expected number of e-folds which can be larger \cite{Liddle:2003as}. The kinetic regime if long enough may affect the tilt of the gravitational waves giving more power to small scales. These effects are much milder if a preheating takes place or if there is minimum in the low energy effective potential. Finally, the supergravity realization of this inflationary scenario may lead to interesting gauge field generation effects. These issues are important and deserve a separate study in order to enable a definite observational distinction between the quadratic chaotic inflation and the exponential potential for an inflaton with kinetic coupling. 

\section{Correspondence between potentials for inflaton with kinetic and minimal coupling}

In the finale, let us depart from the exponential potential example and comment on the general case of monomial potentials
\begin{equation}
V(\phi)=m^{-n+4}\phi^n\,.
\end{equation}
According to the transformation (\ref{trans}), $\varphi=\int g'(\phi)d\phi$, the field $\varphi$  reads 
\begin{equation}
\varphi = \frac{2}{n+2} \frac{\phi^{n/2+1}}{m^{n/2-2}M_P\tilde{M}}\,,
\end{equation}
and it appears to be minimally coupled to gravity during the inflationary phase. Its evolution is governed by the potential $V_m(\varphi)=V[g^{-1}(\varphi)]$
where
\begin{equation} \label{vmm}
V_m(\varphi) = m^{-n+4}\left(\frac{n+2}{2} m^{n/2-2}M_P\tilde{M}\,\varphi \right)^{2n/(n+2)}\,.
\end{equation}
During slow-roll there is direct correspondence between the potential $V(\phi)$ for the kinetically coupled inflaton and the $V_m(\varphi)$ for the minimally coupled inflaton:
\begin{equation} \label{arr}
V\propto \phi^n \quad \longleftrightarrow \quad V_m\propto \varphi^{\frac{2n}{n+2}}\,.
\end{equation}
Let us look into specific examples, starting from the quartic Higgs-like potential. We find that 
\begin{equation}
V(\phi)=\lambda_q\phi^4\quad \longleftrightarrow \quad V_m (\varphi)=\lambda_q^{1/3}(3M_P\tilde{M})^{4/3} \varphi^{4/3}\,,
\end{equation}
i.e. the quartic potential for an inflaton with kinetic coupling \cite{Germani:2010gm} is equivalent to $\varphi^{4/3}$ monomial potential for an inflaton with minimal coupling \cite{Ferrara:2014fqa}.
Also, the quadratic potential $V\propto \phi^2$ with kinetic coupling corresponds to a linear potential $V_m\propto \varphi$,  and, the linear $V\propto \phi$ with kinetic coupling to the $V_m\propto \varphi^{2/3}$;  see also Ref. \cite{McAllister:2014mpa} for relevant monomial potentials in stringy and \cite{Ferrara:2014fqa} in supergravity set ups.

We comment that for the case $n=-2$ the potential expression (\ref{vmm}) cannot be used. The inverse quadratic potential $V(\phi)=m^6\phi^{-2}$ for positive field values is instead depicted to an exponential potential:
\begin{equation}
V(\phi)=\frac{m^6}{\phi^2} \quad \longleftrightarrow \quad V_m(\varphi) = V_{m}(0) e^{-2\frac{M_P\tilde{M}}{m^3}\varphi}\,.
\end{equation}

We also observe that for any positive power $n$ of the monomial potential $V\propto \phi^n$ the corresponding minimal case potential cannot have a power greater than  two. In other words the quadratic potential, $V_m\propto \varphi^2$ is the steepest monomial potential that the $V\propto \phi^n$, $n>0$ can be transformed into, see relation (\ref{arr}), regardless that it may be $n\gg1$. The exponential potential can be seen as the limiting case. Indeed, we find that
\begin{equation}
V(\phi)\propto e^{-\phi}  \quad \longleftrightarrow \quad V_m(\varphi) \propto \varphi^2
\end{equation}
which has been the potential investigated in the previous sections.

\section{Conclusions}

In this Letter we discussed the exponential potentials as candidates for describing the early universe inflationary phase. We found that when the inflaton scalar field is kinetically coupled to the Einstein tensor the exponential potential predictions fit very well the data. The model predicts tensor-to-scalr ratio $r=0.16$ for  $n_s=0.96$ and  $r_* \gtrsim 0.135$ for $N_* \lesssim 60$ e-folds. We found no significant running of the spectral index. These results coincide with the predictions of the quadratic $V\propto \varphi^2$ potential. The underlying reason for this coincidence is that there exists a simple mapping of the one model to the other.  Furthermore, inflation with exponential potential naturally terminates even in the absence of a minimum. This may lead to specific observational signatures that distinguish this model from the quadratic inflationary potential. We also found that this model can be successfully described in a supergravity framework. This allows microscopic theories to accommodate and theoretically motivate this inflationary candidate.

\vskip.1cm

We would like to thank C. Germani  and E. N. Saridakis for discussion and A. Kehagias for insightful comments. 
This work is supported by the Grant agency of the Czech republic under the grant P201/12/G028.

\end{document}